\documentclass[footinbib,a4paper,aps,superscriptaddress,reprint,twocolumn,preprintnumbers,amsmath,amssymb,nobalancelastpage,10pt]{revtex4-1}
\usepackage{color} 
\usepackage{bm}
\usepackage{dcolumn}
\usepackage{times}
\usepackage{amssymb} 
\usepackage{amsmath} 
\usepackage{ragged2e}
\usepackage{graphicx}
\usepackage{epstopdf}
\usepackage[english]{babel}
\usepackage{mathrsfs}
\usepackage{textcomp}
\usepackage{amsthm}
\usepackage{amsfonts}
\usepackage{soul}
\usepackage{braket} 
\usepackage[colorlinks=true,citecolor=blue,linkcolor=blue,urlcolor=blue]{hyperref}
\usepackage{cleveref}
\newcommand{\be}{\begin{equation}}
\newcommand{\ee}{\end{equation}}
\newcommand{\bea}{\begin{eqnarray}}
\newcommand{\eea}{\end{eqnarray}}
\newcommand{\ba}{\begin{array}}
\newcommand{\ea}{\end{array}}

\newcommand{\comment}[1]{{\color{black}#1}}

\begin{document}

\title{Topological quantum critical points in the extended Bose-Hubbard model}

\author{Joana Fraxanet}
\affiliation{ICFO - Institut de Ci\`encies Fot\`oniques, The Barcelona Institute of Science and Technology, 08860 Castelldefels (Barcelona), Spain}
\author{Daniel Gonz\'alez-Cuadra}
\affiliation{ICFO - Institut de Ci\`encies Fot\`oniques, The Barcelona Institute of Science and Technology, 08860 Castelldefels (Barcelona), Spain}
\affiliation{Center for Quantum Physics, University of Innsbruck, 6020 Innsbruck, Austria}
\affiliation{Institute for Quantum Optics and Quantum Information of the Austrian Academy of Sciences, 6020 Innsbruck, Austria}
\author{Tilman Pfau}
\affiliation{5. Physikalisches Institut and Center for Integrated Quantum Science and Technology, Universit\"at Stuttgart, Pfaffenwaldring 57, 70569 Stuttgart, Germany}
\author{Maciej Lewenstein}
\affiliation{ICFO - Institut de Ci\`encies Fot\`oniques, The Barcelona Institute of Science and Technology, 08860 Castelldefels (Barcelona), Spain}
\affiliation{ICREA, Passeig Lluis Companys 23, ES-08010 Barcelona, Spain}
\author{Tim Langen}
\affiliation{5. Physikalisches Institut and Center for Integrated Quantum Science and Technology, Universit\"at Stuttgart, Pfaffenwaldring 57, 70569 Stuttgart, Germany}
\author{Luca Barbiero}
\affiliation{ICFO - Institut de Ci\`encies Fot\`oniques, The Barcelona Institute of Science and Technology, 08860 Castelldefels (Barcelona), Spain}
\affiliation{Institute for Condensed Matter Physics and Complex Systems,DISAT, Politecnico di Torino, I-10129 Torino, Italy}
 
\date{\today}

\begin{abstract}

The combination of topology and quantum criticality can give rise to an exotic mix of counterintuitive effects. Here, we show that unexpected topological properties take place in a paradigmatic strongly-correlated Hamiltonian: the 1D extended Bose-Hubbard model. In particular, we reveal the presence of two distinct topological quantum critical points with localized edge states and gapless bulk excitations. Our results show that the topological critical points separate two phases, one topologically protected and the other topologically trivial, both characterized by a long-range ordered string correlation function. The long-range order persists also at the topological critical points and explains the presence of localized edge states protected by a finite charge gap. Finally, we introduce a super-resolution quantum gas microscopy scheme for dipolar dysprosium atoms, which provides a reliable route towards the experimental study of topological quantum critical points.

\end{abstract}

\maketitle

\paragraph*{Introduction}
Topology represents a fascinating topic in different areas of modern quantum physics \cite{asorey,goldman,ozawa}. Among the related interesting phenomena one finds conducting edge modes in bulk gapped phases \cite{qi,hasan}, quantized conductance \cite{von,laughlin,thouless,moore} and fractionalized charges \cite{tsui,laughlin1,su}. Due to the gapped and non-local nature of their elementary excitations, topological phases further represent a particularly suitable platform where fault-tolerant quantum computation could be performed \cite{kitaev,freedman,bravyi,freedman1,dennis}.\\ 
While in non-interacting systems \cite{altland} symmetry arguments allow for a complete description of topological insulators \cite{kane,bernevig,bernevig1} and superconductors \cite{ryu}, topology in many-body quantum systems represents a more challenging task \cite{kitaev2}. In this context, the notion of symmetry-protected topological (SPT) phases~\cite{pollmann,gu,chen,montorsi} has allowed to classify quantum states that are topologically protected up to local perturbations preserving specific symmetries. Celebrated examples of such fully gapped topological states include the Haldane phase occurring in several strongly correlated systems \cite{haldane,affleck,dalla, berg,nonne,dalmonte,deng,furukawa,rossini,koba,torre13,barbiero, ejima,fazzini,barbiero1,kottmann, hetenyi20, kottmann21}, and the topological Mott insulator (TMI) appearing in interacting Su-Schrieffer-Heger (SSH) models  \cite{manmana,sirker,wang,barbiero2,sbierski,Gonzalez-Cuadra_2019a,Gonzalez-Cuadra_2019b,yu,chanda2021}. Noticeably, the impressive the level of flexibility and accuracy reachable in ultracold atomic platforms has allowed for the experimental realizations of these SPT phases \cite{sompet,leseleuc}.\\ Recent works have further shown that, contrary to standard intuition, topology can also occur in critical phases
\comment{\cite{kestner,fidkowski11,sau11,cheng11, keselman,scaffidi,ruhman17,parker,zhang18,ruben,ruben2,fazzini2,fazzini3,rakovsky20,montorsi,kumar21}}. Here, topological phase transitions deserve special attention. As shown in specific models~\cite{ruben2}, topologically protected edge states (ESs) can remain localized at critical points despite the presence of gapless bulk excitations. Due to the novelty of such a concept, the mechanism that topologically protects quantum critical points is not yet fully understood.\\The results reported in this paper provide a ground for a better comprehension of topological quantum critical points (TQCPs). Our analysis, based on density-matrix renormalization-group (DMRG)~\cite{dmrg,dmrg1,tenpy} calculations, shows that these states of matter occur in distinct regimes of the broadly investigated 1D extended Bose-Hubbard model (EBHM). Here, we unveil that, for large enough interaction, different TQCPs take place both at unit density and at half filling in the presence of a lattice dimerization. Despite the usual vanishing bulk gap at quantum criticality, our results show that topology remains protected by a finite charge gap. This scenario holds in quantum critical points separating a topological and a trivial phase, both sharing a similar type of long-range order captured by the same string correlation function. This hypothesis is confirmed by studying other topological phase transitions between phases without such order, where the charge gap vanishes and the topological properties disappear.\\
We further present a reliable route to probe the existence of TQCPs through a quantum-simulation scheme involving magnetic atoms with dipolar long-range interaction \cite{pfau_rev}. The latter have been recently employed to study droplets physics \cite{schmitt,chomaz1}, thermalization \cite{tang}, supersolidity \cite{guo,tanzi,chomaz}, topological pumping \cite{kao} and, notably, to simulate the EBHM \cite{ferlaino}. However, the dipolar interaction in current experimental setups has been too weak to explore the full EBHM phase diagram. We propose a new quantum gas microscopy scheme based on strongly-interacting dysprosium atoms trapped in a short wavelength optical lattice, which allows us to make precise predictions about the experimental realization and detection of TQCPs. 
\paragraph*{Model}
The 1D EBHM reads as
 \begin{eqnarray}\label{eq:hamiltonian}
 H&=&-\sum_{i=1}^{L-1}(J+\delta J(-1)^i) (b_i^\dagger b_{i+1} + \text{h.c.})+ \nonumber \\
 &&+\frac{U}{2}\sum^{L}_{i=1} n_i(n_i-1)+V\sum_{i=1}^{L-1} n_i n_{i+1},
\end{eqnarray} 
where $b^\dagger_i (b_i)$ is the bosonic operator describing the creation (annihilation) of a particle at the $i$-site of a lattice of length $L$ filled with $N$ bosons. The tunneling amplitude $J-\delta J$ $(J+\delta J)$ describes hopping processes in odd (even) links connecting two nearest-neighbor (NN) sites while $U$ and $V$ represent, respectively, the onsite and NN interactions. The most investigated version of eq.~\eqref{eq:hamiltonian} is the homogeneous case $\delta J=0$ at unit density $\bar{n}=N/L=1$. Here, early studies \cite{monien,monien1} discovered the presence of a gapless superfluid in the limit of weak interaction, a Mott insulator (MI) in the regime of strong onsite repulsion and, for large enough $V$, a charge density wave (CDW$_{1}$). \comment{The CDW$_1$ is characterized by a staggered density pattern of alternated pairs and empty sites, analogous to the effective antiferromagnetic (AF) order of the XXZ spin-1 chain which, for large $U/J$, represents an accurate description of the EBHM \cite{dalla,ejima,deng}. This analogy has induced the discovery of a fully gapped SPT phase with non-local AF order between pairs and empty sites occurring at intermediate couplings strength \cite{dalla}. The latter corresponds to the celebrated Haldane insulator (HI), 
which in the EBHM is protected by effective spin rotational symmetries and exact inversion symmetry ~\cite{ejima_2}.}  Instead, the dimerized case, $\delta J>0$, has been only partially investigated \cite{fabian,sugimoto}. At $\bar{n}=1/2$, the large $U/J$ regime unveils the presence of a phase transition between an \comment{inversion symmetry protected} topological Mott insulator (TMI) and a trivial charge density wave (CDW$_{1/2}$) with local AF order signaled by a staggered density pattern of alternated empty and single occupied sites~\footnote{Similarly to the case of the Haldane phase, inversion symmetry is not enough to protect localized ES siDMRG calculations of the~\cite{fabian}. However, for sufficiently strong on-site interactions, the model is effectively described by its hardcore version, where an extra chiral symmetry guarantees the presence of edge states. The regimes considered here are close to this limit.}. Although the gapped topological phases in eq.~\eqref{eq:hamiltonian} are understood, the study of possible topological states appearing at quantum critical points remains an open challenge.
\paragraph*{Homogeneous case $(\delta J=0)$ at $\bar{n}=1$.}
\begin{figure}
\includegraphics[scale=0.2]{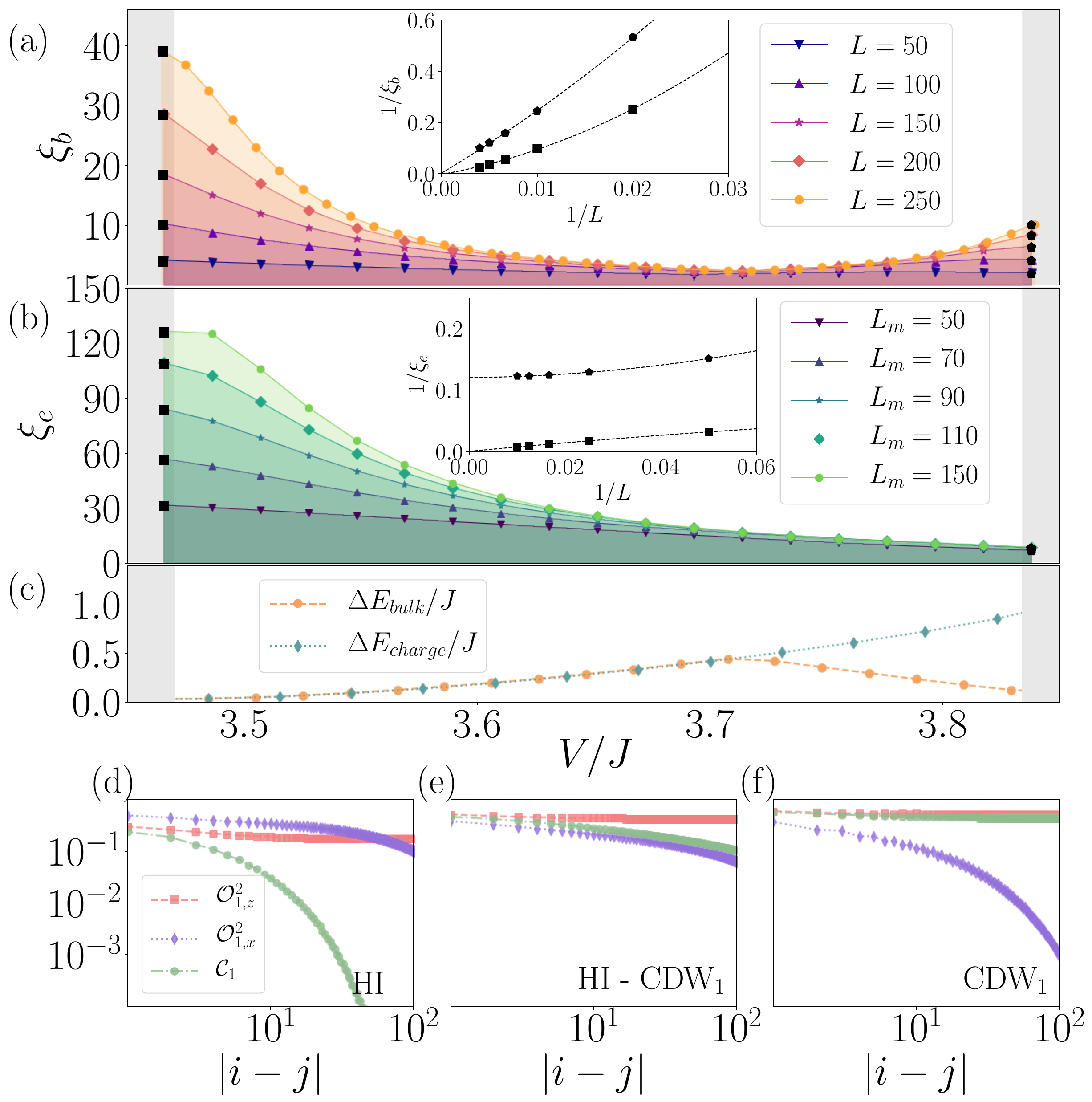}
\caption{\label{fig1} DMRG calculations of edge $\xi_e$ \textit{(a)} and bulk $\xi_b$ \textit{(b)} correlation length in units of the lattice spacing $a_{lat}$, with parameters $\bar{n}=1$, $\delta J=0$, $U/J = 6$ as a function of $V/J$. The insets show the finite size extrapolation of $\xi_e$ and $\xi_b$ at the the MI-HI (squares) and HI-CDW$_1$ (pentagons) critical points. $L_m$ is the maximum length to extract $\xi_e$ from a linear fit of $\log(\vert E_{+}-E_{-}\vert)$ versus $L$ \cite{footnote_plot}. \textit{(c)} Bulk gap $\Delta E_{bulk}$ (orange) and charge gap $\Delta E_{charge}$ (blue) for $L=200$. \comment{The gaps are computed by fixing the edge occupation by means of large chemical potential.} iDMRG calculations of the decay of ${\cal{C}}_1$ (green), ${\cal{O}}^2_{1,x}$ (purple) and ${\cal{O}}^2_{1,z}$ (magenta) relative to: HI at $V/J=3.65$ \textit{(d)}, the HI-CDW$_1$ critical point at $V/J= 3.86$ \textit{(e)} and CDW$_1$ at $V/J= 3.91$ \textit{(f)}. We employ a bond dimension $D=250$ and we cut the number of bosons per site at $n_0 = 4$.}
\end{figure}
SPT phases are characterized by two main features: degenerate localized ESs and gapped bulk excitations. The first requirement is tested by examining the edge localization length $\xi_{e}$, which\comment{, for a phase hosting ESs,} is extracted from $\vert E_{+} - E_{-}\vert\sim e^{-L/ \xi_{e}}$. Here, $E_{\pm}$ are the energies of the two degenerate ground states, $\ket{L}\pm\ket{R}$, where $\ket{L}$($\ket{R}$) denotes a state with the left (right) ES occupied by a bosonic pair and the right (left) ES empty~\cite{NOTE2}. In particular, a thermodynamic extrapolation is expected to show $\xi^{-1}_e\neq0$ in the presence of localized ESs and $\xi^{-1}_{e}\rightarrow 0$ as they delocalize and become bulk states. Analogously, gapped bulk excitations are captured by a finite value of the bulk correlation length $\xi_b\sim\Delta E^{-1}_{bulk}$, where $\Delta E_{bulk}=E_1-E_{GS}$ is the lowest energy bulk gap and $E_{GS}$ ($E_1$) is the energy of the ground (first excited) state. As usual, at quantum criticality $\Delta E_{bulk}=0$ and therefore $\xi_b$ diverges. In Fig.~\ref{fig1}(a-b), our calculations show that the HI has all the aforementioned features, namely $\xi^{-1}_{e},\xi^{-1}_b\neq0$. Probing the SPT nature of HI demands uniquely non-local order parameters. More specifically, the parameters describing the HI are
 ${\cal{O}}_{1,\alpha}(j)=e^{\imath\pi\sum_{k< j}S_{1}^{\alpha}(k)}S_{1}^{\alpha}(j)$, with $\alpha=x,z$, $S_{1}^{x}(i) = \frac{1}{\sqrt{2}}\left(\sqrt{1 - \frac{n_i}{2}}b_i + b_i^\dagger\sqrt{1 - \frac{n_i}{2}}\right)$ and $S^z_{1}(i)=1-n_i$ \cite{deng}. The HI is therefore characterized by the long-range order \comment{\cite{kennedy1992, kennedy1992b}} of the string correlators 
\begin{equation}
{\cal{O}}^2_{1,\alpha}(|i-j|)=\langle{\cal{O}}_{1,\alpha}(i){\cal{O}}_{1,\alpha}(j)\rangle, \qquad \alpha = x, z.\label{oxz}
\end{equation}
Due to the breaking of the effective rotational $x\leftrightarrow z$ symmetry, the two strings ${\cal{O}}^2_{1,\alpha}$ are expected to behave differently in CDW$_1$. In particular, the local AF order of CDW$_{1}$ is characterized by an exponential decay of ${\cal{O}}^2_{1,x}$ but preserved long-range order of  ${\cal{O}}^2_{1,z}$. Crucially, the CDW$_1$ has no topological features since, as discussed, its local magnetic order is captured by the two points correlator
\begin{equation}
{\cal{C}}_1(|i-j|) = \langle S^z_{1}(i)S^z_{1}(j)\rangle\label{cdw}.
\end{equation}
In Figs.~\ref{fig1}(d) and (f) we obtain the expected behavior of eqs. (\ref{oxz}) and (\ref{cdw}) in the HI and the CDW$_1$. By varying $V/J$ at intermediate values of $U/J$, a Gaussian and an Ising phase transition describing respectively the MI-HI and HI-CDW$_1$ critical points are found~\cite{SM}. As expected, and confirmed by the diverging $\xi_b$ shown in Fig.~\ref{fig1}(a), the bulk gap vanishes at the transition points, see Fig.~\ref{fig1}(c). The absence of a bulk gap may suggest that localized ESs also disappear. On the contrary, our calculations demonstrate that this is not always the case. In particular, while at the Gaussian phase transition $\xi^{-1}_b,\xi^{-1}_e=0$, at the HI-CDW$_1$ critical point the edge localization length $\xi_e$ remains finite in the thermodynamic limit. This proves the presence of localized ESs formed by an empty site and a pair of bosons. Moreover, as reported in Fig.~\ref{fig1}(e), this critical point shows algebraic decay of both ${\cal{O}}^2_{1,x}$ and ${\cal{C}}_1$\comment{, while ${\cal{O}}^2_{1,z}$ preserves the long-range order. As discussed, the algebraic decay of ${\cal{C}}_1$ does not occur in the HI, thus implying that a different topological phase exists at this critical point.} \\ In order to understand the origin of such TQCP, we show in Fig.~\ref{fig1}(c) how a gap at energies higher than $\Delta E_{bulk}$, \textit{i.e.} the charge gap $\Delta  E_{charge}=E_{GS}(N+1,L)+E_{GS}(N-1,L)-2E_{GS}(N,L)$ associated to the AF ordering in both HI and CDW$_1$, does not vanish at the HI-CDW$_1$ transition point. This explains why at this TQCP, see Fig.~\ref{fig1} (e),  we still find long-range order of uniquely a string correlator, namely ${\cal{O}}^2_{1,z}$, as required in SPT phases. As a consequence, it is natural to state that the $\Delta E_{charge}$ is the responsible for the protection of the ESs. It is worth to underline that this analysis can be used to interpret the physics of spin-1 chains where the same TQCP takes place \cite{ruben2}. In the latter case, a finite $\Delta E_{charge}$ corresponds to a finite spin gap \cite{ejima} and the string order parameter along the $z-$axis is still expected to be finite at criticality\comment{, see \cite{SM}}. As known ~\cite{deng13,dalla,berg,ejima}, for $U=0$ and truncated local Hilbert space to $n_0=2$, the homogeneous case is strictly equivalent to a spin-1 XXZ chain \cite{note}. Interestingly, our results fully characterize a TQCP away from this spin-1 limit, thus generalizing the physics of TCPQs to new regimes of the EBHM. \comment{We show that for relatively large $U/J$, inversion symmetry combined with effective rotational symmetries are still able to give rise to TQCPs.} We now explore whether similar TQCPs appear when effective spin-1 representations are not applicable.
\begin{figure}
\includegraphics[scale=0.2]{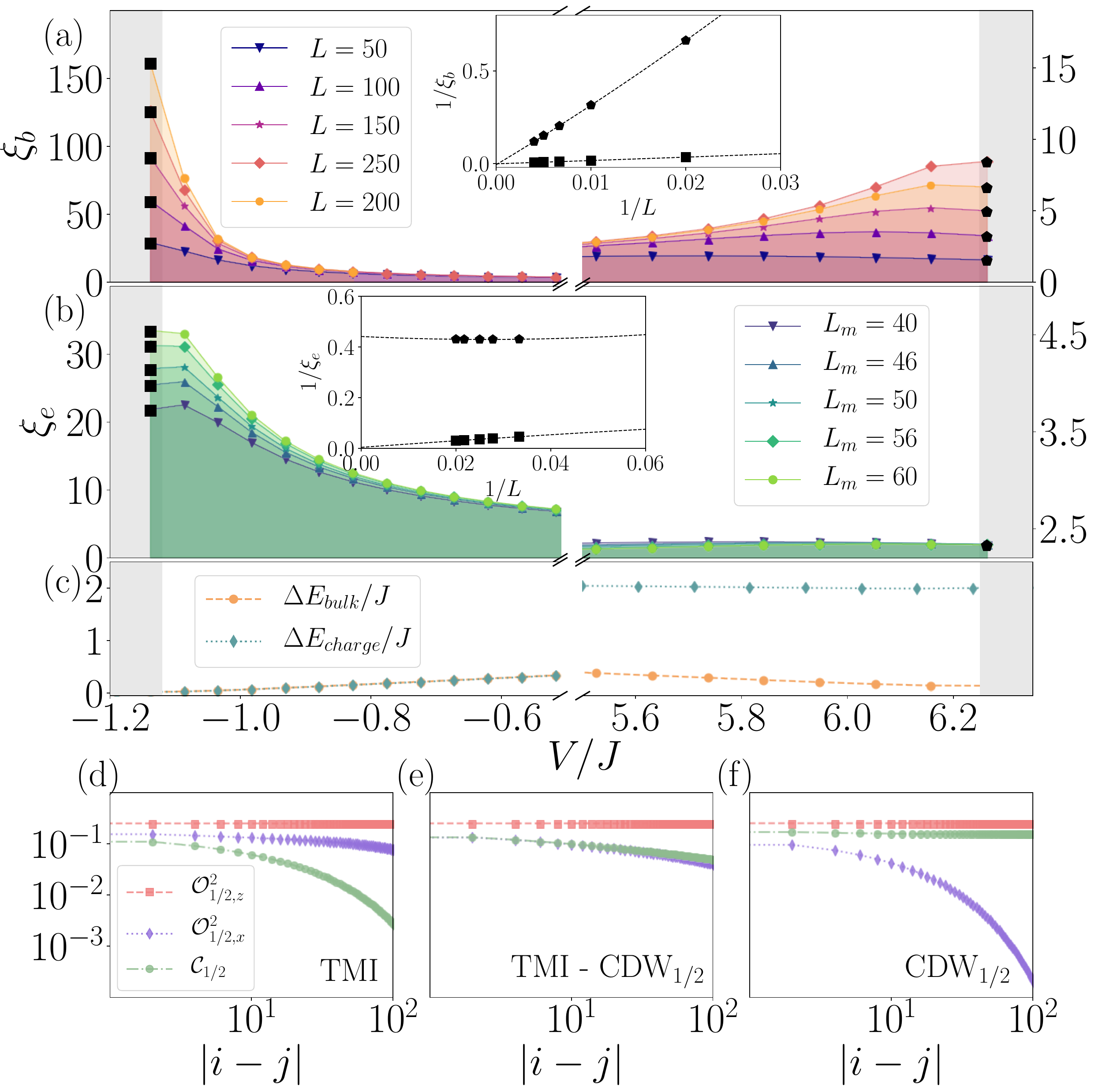}
\caption{\label{fig2} DMRG calculations of the edge $\xi_e$ \textit{(a)} and bulk $\xi_b$ \textit{(b)} correlation length in units of the lattice spacing $a_{lat}$, with parameters $\bar{n}=1/2$, $\delta J=0.5$, $U/J = 6$ as a function of $V/J$. The insets show the finite size extrapolation of $\xi_e$ and $\xi_b$ at the PS-TMI (squares) and TMI-CDW$_{1/2}$ (pentagons) critical points. $L_m$ is the maximum length to extract $\xi_e$ from a linear fit of $\log(E_{+}-E_{-})$ versus $L$ \cite{footnote_plot}. \textit{(c)} Bulk gap $\Delta E_{bulk}$ (orange) and charge gap $\Delta E_{charge}$ (blue) for $L=200$. \comment{The gaps are computed by fixing the edge occupation by means of large chemical potential.} iDMRG calculations of the decay of ${\cal{C}}_{1/2}$ (green), ${\cal{O}}^2_{1/2,x}$ (purple) and ${\cal{O}}^2_{1/2,z}$ (magenta) relative to: TMI at $V/J= 6.03$ \textit{(d)}, the TMI-CDW$_{1/2}$ critical point at $V/J= 6.35$ \textit{(e)} and CDW$_{1/2}$ at $V/J= 6.97$ \textit{(f)}. We employ a bond dimension $D=250$ and we cut the number of bosons per site at $n_0 = 3$.}
\end{figure}
\paragraph*{Dimerized case $(\delta J>0)$ at $\bar{n}=1/2$.}
In this regime, a TMI appears for large enough $U/J$ and intermediate or vanishing $V/J$~\comment{\cite{sugimoto, fabian}}. Notice that such phase displays features analogous to the SPT phase appearing in the dimerized spin-1/2 Heisenberg model \cite{tao}. \comment{The topology of TMI is protected by the inversion symmetry and, although the bulk-edge correspondence does not hold due to the absence of chiral symmetry, localized ESs appear for large enough values of $U/J$ \cite{fabian}, see Fig.~\ref{fig2}(b).} In this case, the ESs are given by either a vanishing or single occupation. Fig. \ref{fig2}(a) shows a finite bulk correlation length in the TMI phase, and in Fig. \ref{fig2}(d) we report that the topological nature of TMI \comment{is effectively captured by} the long-range order of specific non-local string correlators
\begin{equation}
{\cal{O}}^2_{1/2,\alpha}(|i-j|)=\langle{\cal{O}}_{1/2,\alpha}(2i-1){\cal{O}}_{1/2,\alpha}(2j)\rangle, \qquad \alpha = x, z. \label{doxz}
\end{equation}
signaling non-local magnetic order between odd and even sites with alternated bosonic occupation. Here, the associated non-local order parameters are ${\cal{O}}_{1/2,\alpha}(j)=e^{\imath\pi\sum_{k< j}S_{1/2}^{\alpha}(k)}S_{1/2}^{\alpha}(j)$ with $S^z_{1/2}(i)=(1/2-n_i)$ and $S_{1/2}^{x}(i) = 1/2(\sqrt{1 - n_i}b_i + b_i^\dagger\sqrt{1 - n_i})$. For $V<0$, a diverging compressibility signals the presence of a first order phase transition to a regime of phase separation \cite{SM}. At the transition point we find that $\xi^{-1}_{e},\xi^{-1}_{b}=0$, meaning that the ESs disappear and the topology is destroyed. Instead, for larger $V/J$ the system undergoes a phase transition where the TMI is replaced by a trivial CDW$_{1/2}$, captured by the long-range order of the two points correlator
\begin{equation}
{\cal{C}}_{1/2}(|i-j|) = \langle S^z_{1/2}(i)S^z_{1/2}(j)\rangle.
\end{equation}
As in the previous case, the TMI and CDW$_{1/2}$ have respectively non-local and local AF order which now reflects the alternation between empty and singly occupied sites. In Fig.~\ref{fig2}(a) we report that the critical point connecting the CDW$_{1/2}$ and TMI is, as expected, signaled by $\xi_b^{-1} = 0$, which follows from $\Delta E_{bulk}=0$, see Fig.~\ref{fig2}(c). Crucially, in Fig.~\ref{fig2}(b) we reveal that localized ESs captured by a finite $\xi_e$ remain stable at this transition point. The topological ESs are formed by an empty and single occupied state, thus different from the TQCP observed for $\delta J = 0$. Moreover, this TQCP is further characterized by an algebraic decay of ${\cal{O}}^2_{1/2,x}$ and ${\cal{C}}_{1/2}$ (Fig. \ref{fig2}(e)). \comment{In this case, the TQCP is protected uniquely by inversion symmetry, but as shown in \cite{SM}, the same results are found for the hardcore limit, where the topological ESs are formally protected by the combination of inversion symmetry and chiral symmetry.} Furthermore, as we show Fig.~\ref{fig2}(c), the topological ESs in the TQCP and the long-range order of the $z$-oriented string correlator ${\cal{O}}^2_{1/2,z}$, which effectively captures the topological nature of the critical point, are again protected by a non-vanishing charge gap. \\
The results on the dimerized case provide a new example of a TQCP substantially different from the one found for $\delta J=0$. More importantly, they also confirm the hypothesis that the two studied cases share a common mechanism as the responsible for the appearance of TQCPs. Our results point in the direction that if two gapped phases, one topologically protected and the other trivial, are both characterized by a similar long-range order captured by a string correlator, as a consequence, the two phases are connected by a TQCP.
\paragraph*{Experimental realization and detection of a TQCP.}
In this section we discuss how to observe TQCPs using ultracold dysprosium atoms in an optical lattice. In addition to the usual contact interactions, the large magnetic dipole moment of $10\,\mu_B$, where $\mu_B$ denotes the Bohr magneton, makes such atoms interact highly non-locally through dipole-dipole repulsion. Since such interaction scales as $1/r^3$, it is necessary to trap the atoms in a lattice with short periodicity in order to achieve sizeable NN interactions. A lattice formed using lasers with a wavelength around $360\,$nm is promising for this purpose~\cite{SM}. The corresponding lattice spacing $a_{lat}=180\,$nm results in a NN interaction strength of $V/h\sim 200\,$Hz, where $h$ is Planck's constant. This is an enhancement by a factor of six compared to previous experiments with magnetic atoms~\cite{ferlaino} and a factor of four compared to dipolar molecules~\cite{Yan2013}. As shown in Fig.~\ref{fig3}, and explained in detail in \cite{SM}, by changing the lattice depth $V_0/E_R$ and tuning the onsite interaction between atoms using a Feshbach resonance~\cite{Lucioni2018,Schmidt2021}, it is possible to achieve values of interactions where TQCPs can be explored. Notice that in 1D and for not too large $V$, dipolar interaction can neither produce new phases with respect to NN interactions nor change the nature of the transition points. For this reason, our cut of the dipolar interaction to NN is a reliable approximation. Moreover, strongly-coupled bilayer geometries could be realized by implementing a sub-wavelength optical barrier inside the individual lattice sites of a 1D lattice ~\cite{Lacki2016}. In this way, an effective dimerization $\delta J>0$ can be achieved. Thanks to the strong dipolar interaction, we also estimate a critical temperature required to stabilize the topological phase on the order of tens of nK, well within the reach of current experiments. In such a setting, two-point density-density correlation functions ${\cal{C}}_{1}$ and ${\cal{C}}_{1/2}$, string correlators  ${\cal{O}}^2_{1,z}$ and ${\cal{O}}^2_{1/2,z}$, and edge correlation lengths $\xi_e$ are directly accessible using quantum gas microscopy~\cite{Endres2011,Hilker2017,sompet}. However, the requirement of a short lattice wavelength poses a significant challenge for quantum gas microscopy, as the lattice wavelength is significantly below the Abbe resolution limit for the $421\,$nm imaging transition in dysprosium. As we show in \cite{SM}, this limitation can be overcome by using super-resolution techniques to detect individual atoms below the diffraction limit. This scheme results to be inherently number resolving and can allow to individually image empty sites, as well as higher-order occupations, thus allowing for a complete characterization of TQCPs.  
\begin{figure}
\includegraphics[scale=0.2]{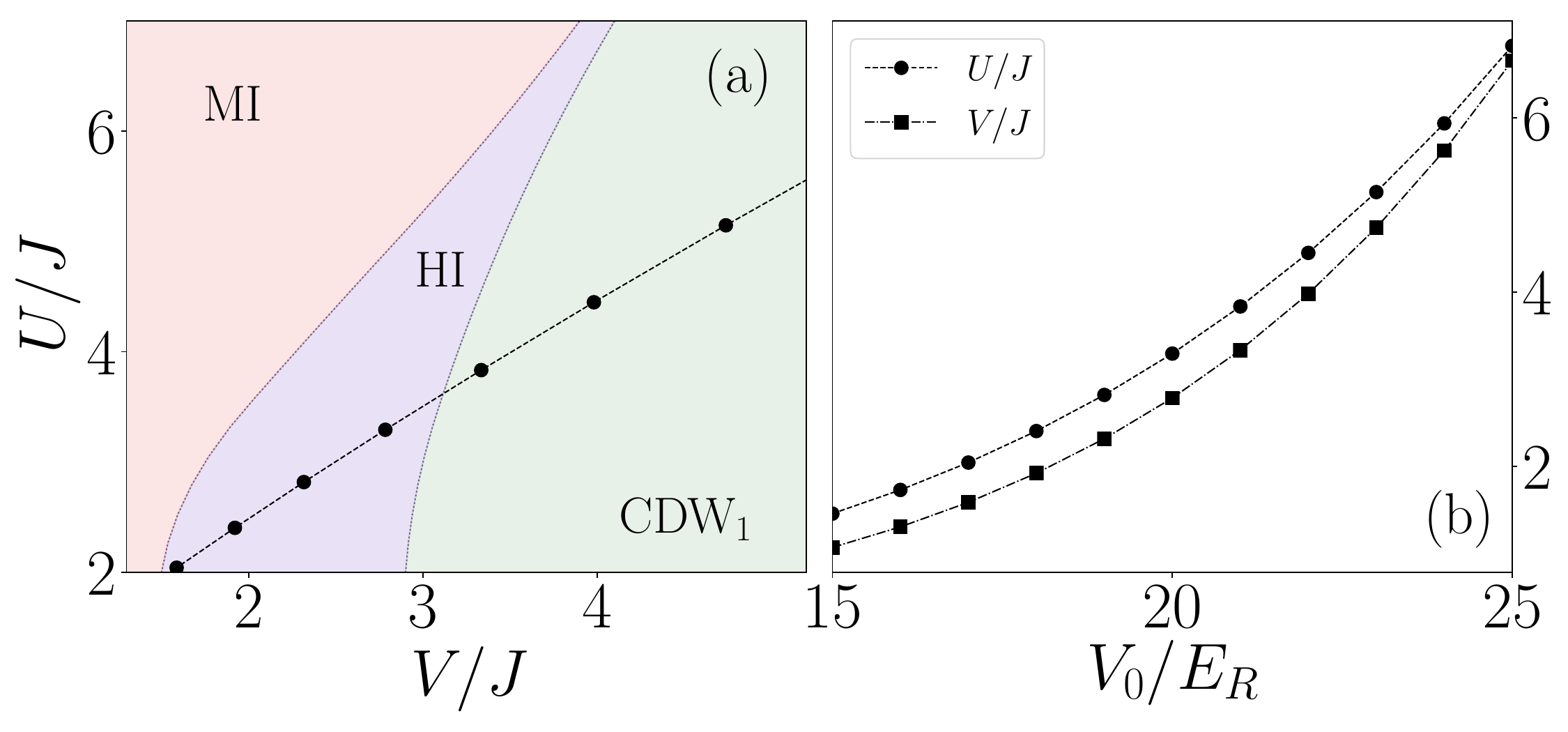}
\caption{\label{fig3} (a) Qualitative phase diagram of eq.~\eqref{eq:hamiltonian} at $\delta J=0$ and $\bar{n} = 1$~\cite{rossini}. The circles represent examples of experimentally accessible values of $U/J$ and $V/J$ calculated by considering dysprosium atoms with dipole moment of $10\,\mu_B$ and $s$-wave scattering length $a_{3D}=91a_0$ with $a_0$ the Bohr radius. The atoms are trapped in a lattice with longitudinal lattice spacing $a_{lat}=180\,$nm and transverse confinement of frequency $\omega_\perp = 2\pi\times 4\,$kHz, see \cite{SM} for details. (b) Values of  $U/J$ (circles) and  $V/J$ (squares) calculated by considering lattice depth $V_0/E_R$ ranging from 15 to 25. Notice that different choices of  $a_{3D}$  and $\omega_\perp$ allow to achieve different sectors of the phase diagram.}
\end{figure}
\paragraph*{Discussion and outlook}
We have shown that the recently discovered phenomenon of topological phases occurring at quantum critical points can be explored in the celebrated EBHM. The TQCPs are SPT phases with gapless bulk excitations but, at the same time, localized edge states. Our analysis has revealed how different kinds of TQCPs are found by adjusting the Hamiltonian parameters. Furthermore, we revealed that TQCPs take place between two fully gapped phases, one topological and one trivial, when they are both characterized by a finite value of the same string correlator denoting a similar type of long-range order. In this scenario, we have demonstrated that, at criticality, a finite charge gap is able to protects the localized edge states. In order to provide a reliable route towards the experimental investigation of TQCPs, we have proposed a detailed experimental setup involving trapped dysprosium atoms at ultracold temperatures. Here, on one hand, laser beams producing short lattice spacing can be employed to achieve the desired region of the phase diagram and, on the other hand, a quantum gas microscope with super-resolution techniques can allow for an accurate detection of TQCPs.\\Due to the fact that EBHMs are of great relevance in different physical systems \cite{bruder,giamarchi, landig, asban, karpov}, we expect that our results can stimulate the study of TQCPs in a broad variety of research lines. 
\nocite{sinha, cremon, cremon2, bartolo, weitenberg11,Wurtz09,McDonald2019, Subhankar2019, Petersen2020, Li2016, lepers, Huang2008}
\paragraph*{Acknowledgments} We thank A. Dauphin, A. Montorsi, G. Palumbo, F. Pollmann and R.Verresen for discussion. J. F.,  D. G-C., M. L. and L. B. acknowledge support from ERC AdG NOQIA, Spanish Ministry MINECO and State Research Agency AEI (Severo Ochoa Center of Excellence CEX2019-000910-S, Plan National FIDEUA PID2019-106901GB-I00/10.13039 / 501100011033, FPI), Fundaci\'o Cellex, Fundaci\'o Mir-Puig, Generalitat de Catalunya (AGAUR Grant No. 2017 SGR 1341, CERCA program, QuantumCAT U16-011424, co-funded by ERDF Operational Program of Catalonia 2014-2020), EU Horizon 2020 FET-OPEN OPTOLogic (Grant No 899794) and the National Science Centre, Poland (Symfonia Grant No. 2016/20/W/ST4/00314), Marie Sk\l odowska-Curie grant STREDCH No 101029393, La Caixa Junior Leaders fellowships (ID100010434), and EU Horizon 2020 under Marie Sk\l odowska-Curie grant agreement No 847648 (LCF/BQ/PI19/11690013, LCF/BQ/PI20/11760031, LCF/BQ/PR20/11770012). D.G.-C. is supported by the Simons Collaboration on Ultra-Quantum Matter, which is a grant from the Simons Foundation (651440, P.Z.). T.L. acknowledges support from the Vector Stiftung and the European Research Council (ERC) under the European Union's Horizon 2020 research and innovation programme (Grant agreement No. 949431). T.P. and T.L. acknowledge support by the German Research Foundation (DFG) within FOR2247 under Pf381/16-1 and Bu2247/1, Pf381/20-1, FUGG INST41/1056-1. We acknowledge QUANT:ERA collaborative project MAQS for financial support.
\bibliographystyle{apsrev4-1}

\begin{thebibliography}{120}
\bibitem [{foo()}]{footnote_plot}%
  \BibitemOpen
  \href@noop {} {}\bibinfo {note} {We extract $-1/\xi_e(L)$ from the slope of
  the linear fit $\log(\vert E_{+}-E_{-}\vert)$ versus $L$, where we take
  values of $L$ from the interval $[L_m-30, L_m]$ (Fig.1) and $[L_m-10, L_m]$
  (Fig.2). We assume that the edge correlation length $\xi_e(L)$ is constant
  within the interval.}\BibitemShut {Stop}%
\bibitem [{NOT()}]{NOTE2}%
  \BibitemOpen
  \href@noop {} {}\bibinfo {note} {In order to select the $\vert L
  \rangle/\vert R \rangle$ configurations for the computation of the edge
  correlation length $\xi_e$, we fix the edge occupation by adding local
  chemical potentials at the edges with opposite sign.}\BibitemShut {Stop}%
\bibitem [{not()}]{note}%
  \BibitemOpen
  \href@noop {} {}\bibinfo {note} {The condition $\bar{n}=1$ translates into
  $S_{tot}^z=0$ when considering spin-1 systems. By using the relation
  $S^z_{i,1}=1-n_i$, one can derive that $\Delta
  E_{charge}=E_{GS}(S_{tot}^z=-1)+E_{GS}(S_{tot}^z=1)-2E_{GS}(S_{tot}^z=0)$
  which is the usual definition of spin gap \cite{chen}.}\BibitemShut {Stop}%
\end{thebibliography}

\end{document}